\documentclass{vldb}

\usepackage[utf8]{inputenc}
\usepackage{blindtext}
\usepackage{amssymb}
\usepackage{graphicx}
\usepackage{xcolor}
\usepackage[british]{babel}
\usepackage{comment}
\usepackage{cite}
\usepackage{url}
\usepackage{breakurl}
\usepackage{balance}
\usepackage{booktabs}
\usepackage{tabularx}
\usepackage{algorithm, algorithmicx, algpseudocode}
\usepackage{pifont}
\usepackage{xspace}
\usepackage{caption}
\usepackage{subcaption}

\makeatletter
\def\@copyrightspace{\relax}
\makeatother

\begin{document}

\title{Towards Global Asset Management in Blockchain Systems}

\author{
\alignauthor
Victor Zakhary \qquad Mohammad Javad Amiri \qquad Sujaya Maiyya\\
Divyakant Agrawal \qquad Amr El Abbadi\\
       \affaddr{Department of Computer Science, University of California Santa Barbara}\\
       \affaddr{Santa Barbara, California} \\
       \email{\{victorzakhary, amiri, sujaya\_maiyya, agrawal, amr\}@cs.ucsb.edu}
}

\maketitle

\begin{abstract}
Permissionless blockchains (e.g., Bitcoin, Ethereum, etc) have shown
a wide success in implementing global scale peer-to-peer cryptocurrency 
systems.
In such blockchains, new currency units are generated through the mining process and are used in addition
to transaction fees to incentivize miners to maintain the blockchain.
Although it is clear how currency units are generated and transacted on, 
it is unclear how to use the infrastructure of permissionless blockchains
to manage other assets than the blockchain's currency units (e.g., cars, houses, etc).
In this paper, we propose
a global asset management system by unifying
permissionless and permissioned blockchains.
A governmental permissioned blockchain authenticates the registration
of end-user assets through smart contract deployments on a permissionless
blockchain. Afterwards, end-users can transact on their assets through
smart contract function calls (e.g., sell a car, rent a room in a house, etc).
In return, end-users get paid in currency units of the same blockchain or 
other blockchains through atomic cross-chain transactions and governmental offices
receive taxes on these transactions in cryptocurrency units.

\end{abstract}

\section{Introduction} \label{sec:introduction}

A blockchain is a distributed data structure for recording transactions
maintained by nodes without a central authority \cite{cachin2017blockchain}.
Nodes in a blockchain system agree on their shared states
across a large network of {\em untrusted} participants.
Existing blockchain systems can be divided into two main categories:
permissionless blockchain systems, e.g., Bitcoin (with PoW-based consensus) \cite{nakamoto2008bitcoin} and
permissioned blockchain systems, e.g., Tendermint  (with BFT-type consensus) \cite{kwon2014tendermint}.

Permissionless blockchains, which are mainly devised for cryptocurrency assets, e.g., Bitcoin \cite{nakamoto2008bitcoin},
are public. Any computing node can participate in maintaining the blockchain without
obtaining a permission from a centralized authority, hence the name permissionless.
In Permissionless blockchains, transactions are used to transfer cryptocurrency assets from one identity to another.
In addition, new currency units are generated through mining; once a new block of transactions is added to the blockchain,
the miner of the block receives some currency units as the mining reward. The amount of mining reward is specified as part of the blockchain protocol.

{\em Permissionless} blockchains are public and computing nodes without a priori 
known identities can join or leave the blockchain network at any time. On the other
hand, a {\em permissioned} blockchain uses a network of a priori known and identified
computing nodes to manage the blockchain. In a permissioned blockchain 
systems, every node maintains a copy of the blockchain ledger and
a consensus protocol is used to ensure that the nodes agree on a unique
order in which entries are appended to the blockchain ledger.
To reach agreement among the nodes, asynchronous fault-tolerant replication protocols have been used.
Nodes in a permissioned blockchain might crash or maliciously behave.
Depending on the failure model of nodes, crash fault-tolerant protocols, e.g., Paxos~\cite{lamport2001paxos},
or Byzantine fault-tolerant protocols, e.g., PBFT~\cite{castro1999practical},
are used to achieve consensus. The tutorial by
C. Mohan~\cite{mohan2017tutorial} provides an overview and 
discusses many aspects of permissioned blockchains.

The blockchain model is similar to an object-oriented programming language (OOPL).
Similar to the primitive data types, user-defined functions, and classes in an OOPL, 
each blockchain also has primitive data types (e.g., an asset, asset ownership, etc) 
and primitive functions operating on these primitive data types (e.g., transactions 
that move currency units from one user identity to another).
Classes and complex functionalities are implemented in the blockchain using 
 smart contracts.
A \textit{\textbf{smart contract}}, as exemplified by Ethereum \cite{wood2014ethereum}, is 
a computer program that self-executes once it is established and deployed.
A smart contract can be seen as a class in an object-oriented programming language
where assets are the objects of that class and transactions update the state (ownership) of the objects.
The state transformation of a smart contract is made persistent in the blockchain
by ensuring that every state change appears as a record in the blockchain.

While permissionless blockchains only support cryptocurrency assets,
smart contracts are more generic and can support any type of asset.
Indeed, a smart contract, like a class in the object-oriented programming, could potentially have different
attributes and functions.
Once a smart contract is written, it can be deployed on a blockchain and
different transactions can call the functions of the smart contract to change its attributes or even
destroy the contract (using a destructor function), making it void.

Deploying general assets (e.g., cars, houses, etc) on the blockchain, 
in contrast to cryptocurrency assets, gives rise to several challenges.
First, ensuring the existence of a registered asset requires some form of 
\textit{authentication} of the asset.
Second, the blockchain system should prevent a malicious end-user
from \textit{double spending} the same asset through two
different smart contracts either 
within the same or on different permissionless blockchains. Finally, depending on 
the asset, the asset transfer should be \textit{legally} allowed by the State law.

To address the aforementioned challenges of
\textit{authentication}, \textit{double spending}, and \textit{legality}
for complex assets in permissionless blockchains, in this paper,
we propose a {\em global asset management system} that unifies permissionless and permissioned blockchains.
In the proposed system, a governmental permissioned blockchain authenticates the registration of end-user assets
through smart contract deployments on a permissionless blockchain.
When an end-user requests to register their assets, in order to prevent \textit{double spending},
a governmental office checks if the asset is not already registered as a smart contract in any
permissionless blockchain.
Next, the governmental office issues
an \textit{authenticated} smart contract registering the asset
wherein the contract also includes the \textit{legal laws} 
associated with the asset and
deploys the smart contract on the permissionless blockchain.
Finally, the end-user will be able  to trade the asset
in the permissionless blockchain while preserving the law enforcement explicitly specified in the smart contract.

Registering complex assets in permissionless blockchains extends the 
transaction model of the permissionless blockchains.
While permissionless blockchains support intra-chain cryptocurrency trades and
cross-chain cryptocurrency trades (with the help of cross-chain swap protocols~\cite{herlihy2018atomic, atomicNolan}),
the proposed system is able to support any type of transactions in either a single or in multiple chains
with any kind of assets.

A key objective of this paper is to demonstrate how global assets can be managed in a blockchain system.
The main contributions of this paper are:

\begin{itemize}
    \item a global asset management system that unifies permissioned and permissionless blockchains
    to manage complex asset,
    \item an extended transaction model that supports varied types of transactions operating on complex assets in multiple blockchains, and finally,
    \item a thorough analysis of the challenges that arise in designing blockchain-based
    asset management systems.
\end{itemize}

The rest of the paper is organized as follows. Section~\ref{sec:permissionless} 
presents the architecture and asset management of permissionless blockchains.
Section~\ref{sec:smart-contracts} explains how smart
contracts are used to extend the functionality of permissionless blockchains.
Section~\ref{sec:permissioned} describes the architecture and asset management of permissioned blockchain.
In Section~\ref{sec:global-asset-management-system}
permissionless and permissioned blockchains are unified
in order to build a novel global asset management system.
We discuss the challenges that arise as a result of this unification in Section~\ref{sec:challenges}.
The related  work is presented in Section~\ref{sec:related-work} and the paper is concluded in 
Section~\ref{sec:conclusion}.
\section{Permissionless Blockchains} \label{sec:permissionless}

Permissionless blockchains are public and therefore, computing nodes, also known 
as miners, can join or leave the blockchain network without obtaining a permission.
Miners maintain a copy of the blockchain ledger and process end-user transactions. Miners and end-users use their public keys as their 
identities in the blockchain system.  Given the open and public model of 
blockchains, these systems are exemplified by the complete absence of the notion of 
trust. That is, these blockchains must operate in spite of a complete absence of any 
trusted entity in the network. Permissionless blockchains are mainly devised for 
cryptocurrency assets, e.g., Bitcoin \cite{nakamoto2008bitcoin}. In this section we 
first explain the architecture of permissionless blockchains and then
present the data and transaction models of such systems.

\subsection{Architecture Overview}

\begin{figure}[ht!]
	\centering
    \includegraphics[width=\columnwidth]{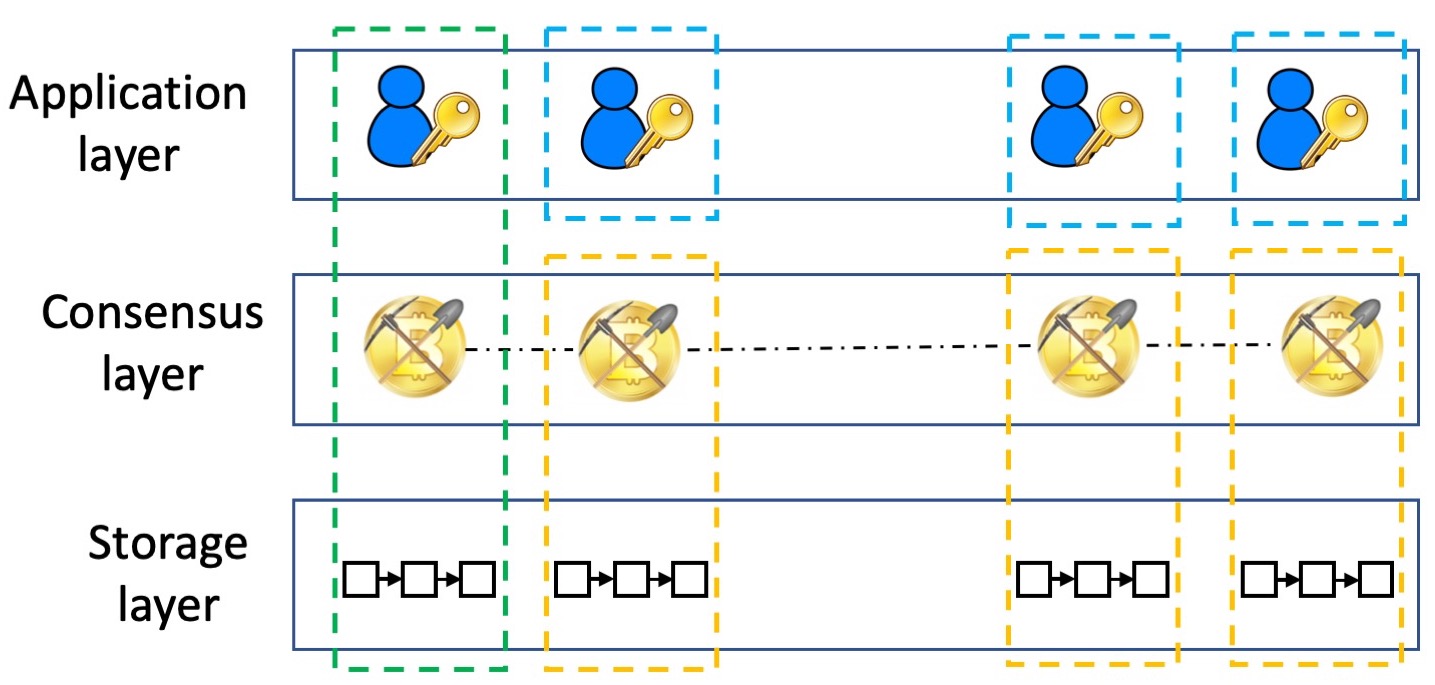}
    \caption{Permissionless Blockchain Architecture Overview}
    \label{fig:permissionlessarch}
\end{figure}

A permissionless blockchain system~\cite{maiyya2018database} 
(e.g., Bitcoin, Ethereum) typically consists of three layers:
an application layer, a consensus layer, and a storage layer, as illustrated in Figure~\ref{fig:permissionlessarch}. 

\medskip
\textbf{The application layer.}
Transactions are initiated by end-users in the application layer.
End-users have identities, defined by their public keys and signatures, generated using their private keys. 
Digital signatures are the end-users' way to generate transactions.
Once transactions are generated, the users multicast their transactions
to the mining nodes in the consensus layer through a client library.
Transactions are used to transfer assets from one end-user identity to another.

\medskip
\textbf{The Consensus Layer.}
In permissionless blockchains consensus is established through mining.
A mining node validates the transactions it receives, puts the valid transactions
into a block and try to solve some cryptographic puzzle.
The industrious miner who solves the puzzle multicasts the block to all nodes.
To make progress, when a miner receives a block of transactions,
it first validates the solution to the puzzle and all transactions in the block,
appends the block to the blockchain, and then proceeds to mine the next block.

\medskip
\textbf{The storage layer.} The ledger is a tamper-proof 
chain of blocks that is maintained by every mining node.
The storage layer comprises a decentralized 
distributed ledger managed by an open network of nodes.
Each block of the ledger contains a set of 
valid transactions that transfer assets among end-users. 

Nodes in a permissionless blockchain are either end-users or miners.
While end-users have only the application layer,
the architecture of miners consist of the consensus and storage layers.
Note that a miner can also be an end-user, thus has all three layers.

\subsection{Asset Management} \label{sub:data_model}

From a data point of view, assets in a permissionless blockchains can be modeled
using data types, i.e., an asset is represented by currency units and its ownership.
Transactions, on the other hand, transfer the ownership of assets,
i.e., move some currency units from one user identity to another user identity.

The ownership information of assets is stored in the storage layer.
The owner of an asset is determined using identities that
are implemented using public keys.
A coin that is linked to a user's public key is owned by that user.

Transactions transfer the ownership of an
asset from one identity to another.
A transaction is basically a digital signature.
End-users, in the application layer,
use their private keys~\cite{rivest1978method} to digitally sign 
assets linked to their identity to transfer these assets to other 
identities, identified by their public keys. These digital signatures
are submitted to the consensus layer via message passing through a 
client library. It is the responsibility of the miners to validate 
that end-users can transact only on their own assets. If an 
end-user digitally signs an asset that is not owned by this end-user,
the resulting transaction is not valid and is rejected by the miners.
In addition, miners validate that an asset cannot be spent twice and hence
prevent double spending of assets.
Using transactions, an asset can be tracked 
from its registration in the blockchain, the first owner, to its latest owner 
in the blockchain.
Transactions are stored in the blockchain in the storage layer.

Registration and divisibility are two other aspects of asset management.
In bitcoin and many other cryptocurrencies, new coins are generated and registered
in the blockchain through mining. In fact, once a miner solves the puzzle, it is allowed to generate some
amount of coin as a mining reward.

Assets can be split or merged using transactions.
Each transaction takes one or more input assets owned by one identity 
and outputs one or more assets where each output asset
is owned by one identity. Indeed, a transaction references previous transaction outputs
as new transaction inputs and dedicates all input coin values to new outputs.
The summation of a transaction's input assets matches the 
summation of its output assets assuming that no transaction fees are 
imposed.

\begin{figure}[ht!]
	\centering    \includegraphics[width=\columnwidth]{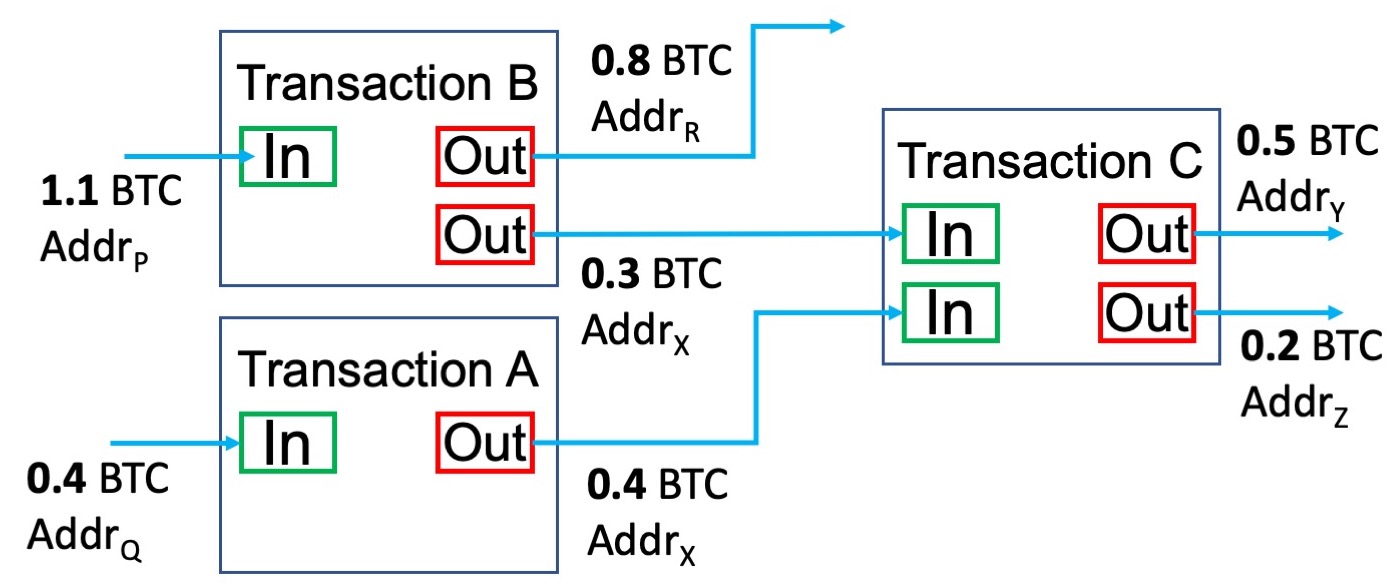}
    \caption{Transactions input and output in blockchain}
    \label{fig:blockchain_transactions}
\end{figure}

Figure~\ref{fig:blockchain_transactions} shows an example of three transactions $A$, $B$, and $C$
in the Bitcoin blockchain. 
As can be seen, in transaction $A$,
a user (with address $addr_Q$) transfers $0.4$ bitcoins to another user $addr_X$.
In Transaction $B$, $addr_P$ splits $1.1$ bitcoins to (1) $0.8$ to $addr_R$ and (2) $0.3$ to $addr_X$.
Finally, in Transaction $C$, the outputs of transactions $A$ and $B$ that are owned by $addr_X$
($0.4$ bitcoins from transaction $A$ and $0.3$ bitcoins from transaction $B$)
are merged and then split to $0.5$ to $addr_Y$ and $0.2$ to $addr_Z$.

In traditional databases, end-user transactions execute arbitrary updates in the storage
layer as long as the semantic and the access control rights of a transaction are validated in the 
application layer.
On the other hand, in blockchain systems, this validation is 
explicitly enforced in the consensus layer and hence end-users, in the 
application layer, are allowed to transact only on the assets they own in the storage layer.
This is in contrast to the database systems model where individual transactions in isolation
and in the absence of concurrency are assumed to be \emph{correct}.
Indeed, the database concurrency control component provides the guarantee that
the interleaved execution of multiple transactions will be equivalent to some serial execution.
In the blockchain context, however, the correctness of individual user transactions
cannot be assumed due to the absence of a trust model and hence
the underlying storage system checks the validity of the user transactions.

Note that this is only feasible due to the restrictive semantics of the currency-based asset model.
More complex applications on permissionless blockchain also need to deal with the lack of assumption
of the \emph{correct transaction model} that is made in traditional database systems.

\section{Smart Contracts} \label{sec:smart-contracts}

Blockchains can be viewed as analogous to object-oriented
programming languages. Consider
permissionless blockchains where each blockchain consists of primitive data types such as an asset 
represented by currency units, user identities, user accounts, etc, along with primitive 
functions that are applicable on these primitive data types such as transactions that
move some currency units from one user identity to another user identity. 
To represent a complex asset, analogous to a complex data type, an end-user 
writes a \textbf{smart contract}~\cite{buterin2014next} that represents this 
complex asset. 
A smart contract is a program written in some scripting language 
(e.g., Solidity for Ethereum smart contracts~\cite{solidity}) that allows general 
program executions on a blockchain's mining nodes. A smart contract can be 
thought of as a class in an object-oriented programming language. End-users write
the specification of the member variables (the state) and the member 
functions (the state transitions) of this class in the smart contract 
code. For example, Alice can write a smart contract that represents her ownership
of a car. The member variables of this smart contract could include
the car attributes (e.g., make, model, year, the VIN), 
the car owner (Alice's public key), and the sell price of the car 
(e.g., $10$ bitcoins). The member functions could include a buy function that allows
Alice to move the ownership of the car to another end-user if this end-user pays
Alice the car price through a buy function call. 

After an end-user writes the description of the smart contract class, the 
end-user deploys the smart contract on a blockchain through a 
\textit{deployment message} that is sent to the mining nodes in the 
consensus layer. The deployment message includes the 
smart contract code. Deploying a smart contract on the blockchain instantiates an 
object~\cite{herlihy2019blockchains,dickerson2017adding} of the smart contract 
class and stores this object in the blockchain.
This object has a state, a constructor that is called when 
a smart contract is first deployed on the blockchain, and a set
of functions that could alter the state of this object. The 
constructor initializes the object state. To alter the state of 
the object, end-users call smart contract functions via \textit{function call 
messages}. End-users send function call messages to the mining nodes 
accompanied by the function parameters to the blockchain mining nodes. Miners 
execute the function on the current contract state and record any contract 
state transitions in their current block in the  blockchain. Therefore, a smart 
contract state might span many blocks after the block where the smart contract is first deployed.
The deployment message is a special case of a function call message
that includes the smart contract code and results in executing the constructor
of this smart contract. End-users pay a fee to the mining nodes for every
function call message, including the deployment message, to incentivize the mining
nodes to execute this function and record the state transitions of the smart
contract object in their current block.

Every function call message, $msg$, includes some implicit parameters that
are passed in the message and are accessible by the function code. These parameters 
include the sender end-user public key, accessed through $msg.sender$, and 
an optional asset value, accessed through $msg.val$. This optional asset value 
allows end-users to use their assets, in currency units, in the smart contract
functions. For example, Alice might deploy a smart
contract that locks $10$ ethers of hers in the contract, passed in the deployment 
message, and conditionally transfers these $10$ ethers to Bob if Bob solves some 
puzzle that is written in the contract. Another example is the car ownership
transfer where Bob passes $10$ bitcoins of his in the buy function call of Alice's
smart contract in order to buy Alice's car. Note that miners have to verify that 
end-users who pass an asset value in a smart contract function call must 
own this asset value and they cannot double spend this asset value
in another smart contract function call or another implicit transaction.

\begin{figure}[ht!]
	\centering
     \includegraphics[width=\columnwidth]{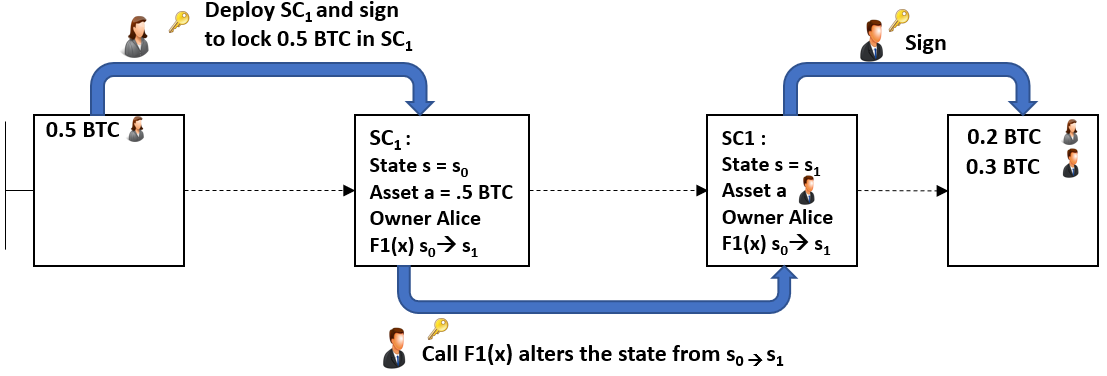}
    \caption{Smart contract state can span multiple blocks in the blockchain.}
    \label{fig:smart-contract-spans}
\end{figure}

Figure~\ref{fig:smart-contract-spans} illustrates a smart contract example
where the smart contract state spans multiple blocks in the blockchain. As
shown, Alice deploys smart contract $SC_1$ on the Bitcoin blockchains. Along
with the deployment message, Alice passes her $0.5$ bitcoins signed to be
locked in $SC_1$. This locking moves the ownership of the $0.5$ bitcoins from
Alice to $SC_1$. $SC_1$ has a state variable $s = s_0$, an asset $a$ ($0.5$ BTC), 
and an owner \textit{Alice}. $SC_1$ has a function $F1(x)$ that transfers 
the ownership of the asset $a$ to any caller who provides a valid parameter 
$x$ according $F1$'s logic. Also, $F1(x)$ transfers $SC_1$ state variable $s$ 
from $s_0$ to $s_1$. When Bob calls $F1(x)$ providing a valid parameter $x$, the 
mining nodes execute this function call and record all the smart
contract state transitions in their current block. As shown, after Bob calls
$F1(x)$, the contract's variable $s$ is set to $s_1$ and the asset $a$, $0.5$ bitcoin,
is moved to Bob. Bob can spend the transferred asset via transactions
in the following blocks. In Figure~\ref{fig:smart-contract-spans}, the ownership of
the $0.5$ bitcoins is moved from Alice to $SC_1$ through the deployment message, 
from $SC_1$ to Bob through the $F1$ function call, and finally is split among 
Alice ($0.2$ BTC) and Bob ($0.3$ BTC) via a Bitcoin transaction.

\begin{algorithm}[t]
\caption{Smart contract that represents a car as a complex asset}
 \label{algo:car-smart-contract}

{\sf class} CarSmartContract \{
    
\begin{algorithmic}[1]
    \State {\sf String} make  \Comment{{\tt the make of the car}}
    \label{line:car-make}
    \State {\sf String} model \Comment{{\tt the model of the car}}
    \State {\sf Integer} year \Comment{{\tt the manufacture year of the car}}
    \State {\sf Double} p     \Comment{{\tt the price of the car}}
    \label{line:car-price}
    \State {\sf Address} o    \Comment{{\tt the public key of the owner}}
    \label{line:car-owner}
    
    \Procedure{Constructor}{{\sf String} make, {\sf String} model, {\sf Integer} year, {\sf Double} p} \label{line:car-constructor}
        \State {\sf this}.make = make
        \State {\sf this}.model = model
        \State {\sf this}.year = year
        \State {\sf this}.p = p
        \State {\sf this}.o = $msg.sender$ \label{line:car-owner-address}
    \EndProcedure
    
    \Procedure{Buy}{Address curOwner} \label{line:car-buy}
        \State {\sf requires}($msg.val \ge $ {\sf this}.p and curOwner == this.o) \label{line:price-requirement}
        \State {\sf transfer} $msg.val$ to {\sf this}.o
        \State {\sf this}.o = $msg.sender$
    \EndProcedure
    
    \Procedure{UpdatePrice}{Double p} \label{line:car-update-price}
        \State {\sf requires}($msg.sender == $ {\sf this}.o)
        \State {\sf this}.p = p
    \EndProcedure
    
\end{algorithmic}
\}
\end{algorithm}

Algorithm~\ref{algo:car-smart-contract} shows an example of a smart contract
to register a car as a complex asset. The member variable (Lines~\ref{line:car-make}
--~\ref{line:car-owner}) represent the attributes of the car. The
constructor (Line~\ref{line:car-constructor}) initializes the car object with the 
attribute values passed in the deployment message (e.g., {\sf make}, {\sf model}, {\sf year}, and 
{\sf price}). In addition, the constructor uses the implicit parameter {\sf msg.sender} to
initialize the car owner (Line~\ref{line:car-owner-address}). The smart contract
has two other functions: {\sf Buy} (Line~\ref{line:car-buy}) and {\sf UpdatePrice} 
(Line~\ref{line:car-update-price}). The {\sf Buy} function allows other end-users to 
buy the car asset. An end-user who wants to buy the car sends a {\sf Buy} function call 
message accompanied by the implicit parameters {\sf msg.sender}  and {\sf msg.val}
in addition to, an explicit parameter {\sf curOwner} that includes the address of the
current car owner. 
{\sf msg.sender} determines the identity of the end-user who wants to buy the car and 
{\sf msg.val} determines the value in currency units that the end-user wants to pay for 
the car. {\sf curOwner} determines the current owner of the asset from the perspective
of the function call request. This is necessary to prevent concurrent Buy requests
from buying the same asset. Assume \textit{two} concurrent Buy function calls that 
are submitted with the same {\sf curOwner} value. If one Buy request succeeds, the owner 
of the asset will be altered as a result. Therefore, the other Buy request will fail. The Buy function requires {\sf msg.val} to be greater than or equal to the car price 
and {\sf curOwner} to be equal to the current car owner (Line~\ref{line:price-requirement}). If true, {\sf msg.val} is transferred to the current 
owner and the ownership of the car is transferred to {\sf msg.sender}. However, if the 
\textit{requires} instruction fails, the function
execution is terminated and the transfers do not take place. Finally,
the function {\sf UpdatePrice} allows only the current owner of the car to update
its price.


Although smart contracts are powerful tools to represent the attributes and the 
functionality of complex assets in permissionless blockchain, registering 
complex assets via smart contract deployments faces several challenges including
{\em the authentication}, {\em double spending}, and {\em legality}.

\medskip
\textbf{The authentication challenge.}
``How can end-users 
{\em authenticate the registered asset} and ensure its existence?". For example, if Alice
registers her car title in the Bitcoin blockchain, ``how could Bob who wants to buy 
this car authenticate that this car physically exists and that Alice
is not maliciously registering a car that does not exist?". 

\medskip
\textbf{The double spending challenge.}
``How can the blockchain system 
prevent a malicious end-user from {\em registering the same asset in two smart contracts} 
within the same or in different permissionless 
blockchains?". In the previous example, even if Bob could magically authenticate 
Alice's car smart contract, "how could Bob ensure that this is the only smart 
contract that Alice deployed to register her car in a permissionless blockchain?". 

\medskip
\textbf{The legality challenge.}
``How can end-users ensure
that this asset transfer is {\em legally allowed by State law} where this transfer
takes place?". This challenge addresses the State laws including the taxation law. 
Transferring the ownership of a car requires the buyer to pay a transfer taxes 
to the State according to the State law. 

Our proposal in Section~\ref{sec:global-asset-management-system} 
addresses these challenges by unifying both permissioned and 
permissionless blockchains. This unification allows end-users to use
the infrastructure of permissionless blockchains to trade
their assets without violating State laws while preventing double 
spending and trading unauthenticated assets.
\section{Permissioned Blockchains} \label{sec:permissioned}

In a blockchain, nodes agree on their shared states across a network of participants.
Existing blockchain systems can be divided into two main categories of
permissionless and permissioned blockchains. While {\em Permissionless} blockchains 
are public and any computing node can participate in maintaining the blockchain 
ledger, {\em 
permissioned} blockchain consists of a set of known and identified nodes that do not 
fully trust each other.

Blockchain was originally devised for Bitcoin cryptocurrency \cite{nakamoto2008bitcoin},
however, recent systems focus on its unique features such as
transparency, provenance, fault-tolerant, and authenticity
to deploy a wide range of distributed applications such as
supply chain management, IoT, and healthcare
in a permissioned settings.

\subsection{Architecture Overview}

The architecture of a permissioned blockchain consists of
{\em Application layer}, {\em Consensus layer}, and {\em Storage layer}.
The application layer of a permissioned blockchain, similar to permissionless blockchains,
consists of end-users who submit their transactions to the blockchain through a client library.
However the consensus layer which is mainly responsible for ordering and validating the transactions differs
from the consensus layer in permissionless blockchains.
In fact, since the nodes in a permissioned blockchain are known and identified,
mining can be replaced with traditional consensus protocols
in order to establish a total order on the requests \cite{cachin2016architecture}.
Finally, the storage layer, similar to permissionless blockchains, consists of a decentralized  distributed ledger
maintained by every node within the blockchain.

\begin{figure}[t]
	\centering
    \includegraphics[width=\columnwidth]{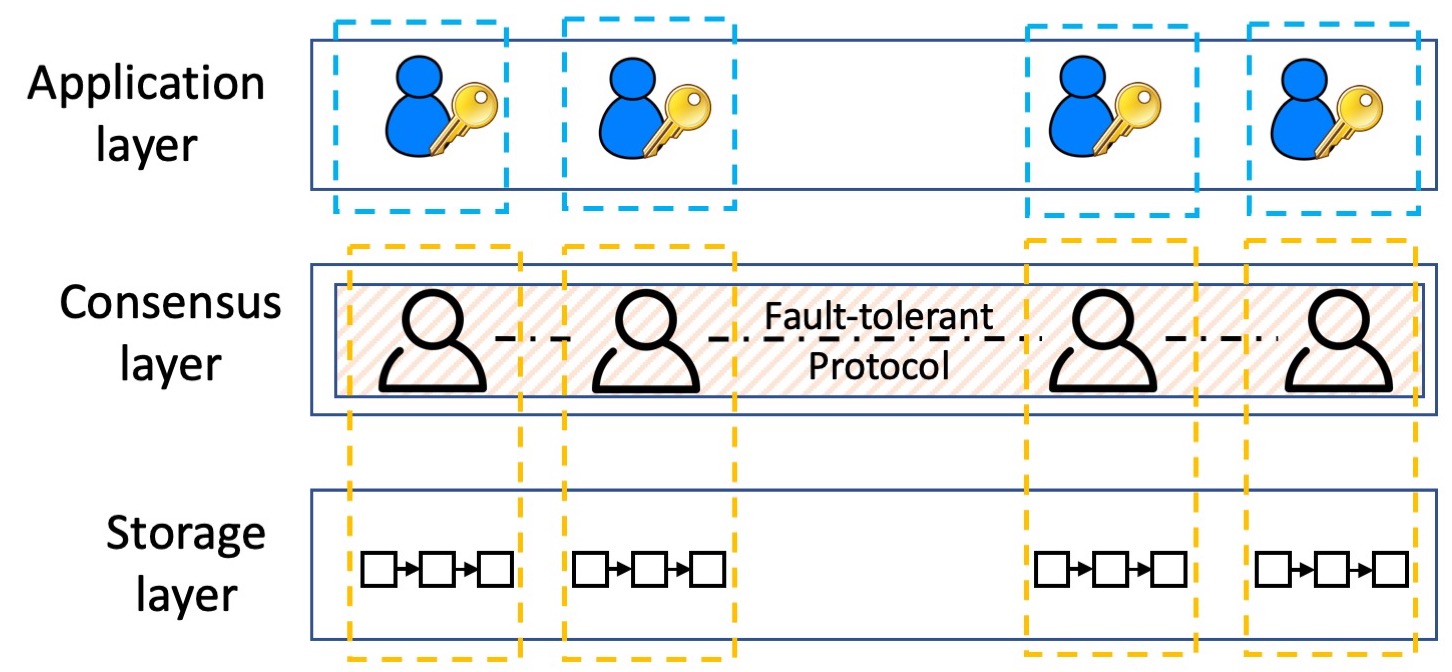}
    \caption{Permissioned Blockchain Architecture Overview}
    \label{fig:architecture}
\end{figure}

The consensus layer runs a consesus protocol among the computing nodes of the 
consensus layer. Consensus protocols employ State Machine Replication (SMR) technique 
to replicates data, e.g. ledger, over nodes.
State machine replication is a technique for implementing a fault-tolerant service
by replicating servers \cite{lamport1978time}.
In the state machine replication model replicas agree on
an ordering of incoming requests.

To establish consensus among the nodes in a permissioned blockchain, asynchronous fault-tolerant protocols can be used.
Nodes in a permissioned blockchain might crash or maliciously behave.
In a crash failure model, nodes operate at arbitrary speed,
may fail by stopping, and may restart, however, they may not collude, lie, or
otherwise, attempt to subvert the protocol.
Whereas, in a Byzantine failure model, faulty nodes may exhibit arbitrary,
potentially malicious, behavior.

Crash fault-tolerant protocols
guarantee safety in an asynchronous network using $2f{+}1$ nodes
to overcome the simultaneous crash failure of any $f$ nodes while
in Byzantine fault-tolerant protocols,
$3f{+}1$ nodes are usually needed to provide the safety property in the presence of $f$ malicious nodes.

Permissioned blockchain mainly follow an order-execute paradigm where
a set of peers (might be all of them) validates the transactions,
agrees on a total order for the transactions,
puts them into blocks and multicasts them to all the nodes.
Each node then validates the block, executes the transactions using a "smart contract", and updates the ledger.


\subsection{Data Management}

The permissioned blockchain systems are distinguished from the permissionless blockchain systems in one critical way:
although there is in general a lack of trust among entities, all entities or components in the system are completely identified. 
The identified storage nodes in the permissioned blockchains can come together to allow a much more general-purpose data model then that is stipulated in the permissionless system. Thus, Permissioned blockchains can be used for different distributed applications.
In the same vein, since end-users of permissioned systems have known identities, we can enforce the \emph{correct transaction computation} assumption from the database systems. This allows the transaction models in permissioned system to be more general than the transaction model in permissionless blockchains
where each transaction mainly transfers the ownership of assets, i.e., cryptocurrencies, from one identity to another.
In a permissioned blockchain depending on the application, different types of transactions can be defined.
For example, a Supply Chain Management includes different processes such as
farming, refining, design, manufacturing, packaging, and transportation.
As a result, to support a Supply Chain Management system, the
permissioned blockchain should be able to record all transactions within these different processes.

To summarize, permissionless systems are completely open and public and therefore do not have the notion of identity and trust. In effect, these systems have to withstand malicious behavior at all levels: at the level of an end-user, at the level of a network node who is miner, as well as other nodes that try to compromise the sanctity of the system. End-users, consensus nodes, and storage nodes in the permissioned system, on the other hand, all have known identities. The lack of trust is primarily because of two possibilities. If the permissioned system belongs to a single enterprise,  the consensus and storage nodes may be stored at different infrastructure providers. The source of maliciousness in this setting may arise if one or more of the infrastructures are compromised. Alternatively, the permissioned system may be a designed to facilitate cooperation among multiple enterprises. The source of maliciousness in this setting may arise due to the competition among these cooperating entities.

\section{Global Asset Management System}\label{sec:global-asset-management-system}

This section proposes a global asset management system that leverages
permissioned blockchains to address the \textit{authentication}, the 
\textit{double spending}, and the \textit{legality} challenges of 
using smart contract to represent complex assets in permissionless 
blockchains. Governmental offices deploy their own 
permissioned blockchains.  End-users request from a governmental
office to register their assets in a smart contract in some 
permissionless blockchain. End-users pay a registration fee to the 
governmental office for this registration. The governmental office 
checks if this asset has not been previously registered in any 
permissionless blockchain smart contract. This check is necessary to 
ensure that end-users cannot \textit{double spend} their 
assets through several smart contracts. If true, the governmental office issues
an \textit{authenticated} smart contract to deploy on the permissionless 
blockchain included in the registration request. The governmental office encodes 
the legal laws, including the taxation law, in the terms of the smart contract. 
Afterwards, the governmental office deploys the smart contract on behalf of the 
end-user. The smart contract, owned by the governmental office identity, registers
an asset, owned by the end-user identity, and allows the end-user to trade the
asset in the permissionless blockchain while preserving the legal rights of
the governmental office. For example, the California DMV office deploys a car 
registration permissioned blockchains. When Alice wants to register her car in 
the Ethereum blockchain, she requests a smart contract registration of her car
in the Ethereum blockchain from the DMV office. The DMV office issues this smart
contract stating that any transfer of ownership of this car should pay the
governmental office some tax percentage, say 10\%, from the car price. Alice cannot
double spend her car as there exists only one smart contract that represents Alice's
car in any permissionless blockchain. Now, if Bob wants to buy Alice's car, Bob first
checks that this smart contract is authenticated by the governmental office identity
to ensure the authenticity of the car in the smart contract. If true, Bob can buy
the car by submitting a Buy function call request to the mining nodes of the
permissionless blockchain. This request is accompanied by Bob's currency units
that he wants to pay for the car in the implicit parameter $msg.val$. If
the Buy function call succeeds, Alice gets paid in currency units, the governmental
office gets paid a tax in currency units, and the ownership of the car is transferred
to Bob.

This proposal simplifies the process of trading assets by leveraging
the permissionless blockchain infrastructure. Once an asset is registered in a smart 
contract, trading
this asset among end-users is as simple as a permissionless blockchain transaction. 
End-users are motivated to register their assets as this registration
offers them an elimination of the bureaucratic process needed to trade their assets. 
Governmental offices are motivated to participate by running a permissioned 
blockchain as it offers them automation and transparency. 

In Section~\ref{sub:proposal-architecture}, we present the architecture overview
of the permissioned and permissionless blockchain unification proposal. Then, 
we explain the transaction model of the registered assets in permissionless
blockchain in Section~\ref{sub:proposed-transaction-model}. We 
present a car registration smart contract example in 
Section~\ref{sub:smart-contract-example}. Finally, we discuss alternative 
asset management models in Section~\ref{sub:alternative-model}.

\subsection{Architecture Overview}\label{sub:proposal-architecture}

\begin{figure}[ht!]
	\centering
     \includegraphics[width=\columnwidth]{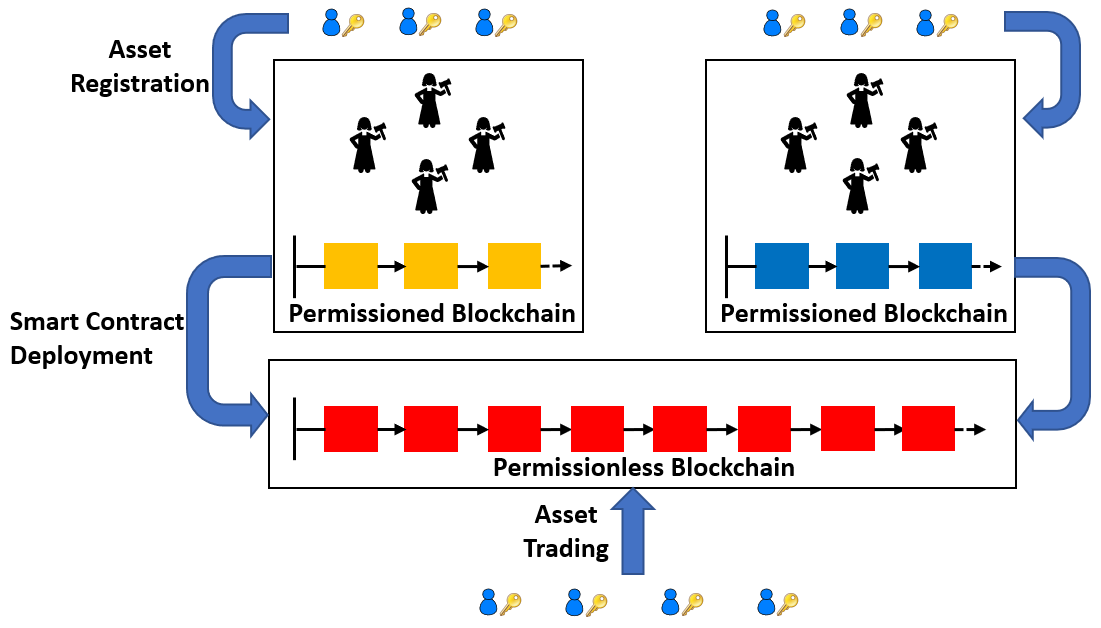}
    \caption{Architecture overview of the permissioned and permissionless blockchain unification proposal.}
    \label{fig:proposed-architecture}
\end{figure}

Figure~\ref{fig:proposed-architecture} illustrates the architecture overview
of the permissioned and permissionless blockchain unification proposal.
Governmental  offices run their trusted asset registration systems.
These trusted asset registration systems could be as simple as a database management systems.
Governmental offices run permissioned blockchains with a set of trusted governmental officials,
called {\em validators}.
Such governmental officials might fail, e.g., the identity of a governmental official gets stolen.
Depending on the failure model of the validators, they run a 
crash fault-tolerant, e.g., Paxos~\cite{lamport2001paxos} or 
a Byzantine fault-tolerant, e.g., PBFT~\cite{castro1999practical} consensus protocol
to agree on the registered assets.

An end-user sends an {\em asset registration} request to the permssioned blockchain validators.
Validators run a consensus protocol among themselves to ensure that this asset is
not previously registered.
Once the consensus is achieved, each validator executes the request using some predetermined smart contract.
The smart contract generates another smart contract representing the registered asset.
To ensure deterministic execution of transactions, as mentioned in Section~\ref{sec:smart-contracts},
smart contracts are written in scripting languages like Solidity.

Validators then add the asset registration record in their permissioned blockchain.
In addition, they authenticate the deployment of the resulted smart contract in a permissionless blockchain.
This  smart contract is owned by a multi-signature address of the validators.
In addition, the asset in the smart contract is owned by the end-user identity.
Once the smart contract is deployed on the permissionless blockchain, end-users can
trade assets through smart contract function calls.

As shown in Figure~\ref{fig:proposed-architecture}, different governmental offices
can use the same permissionless blockchain to deploy their asset registration
smart contracts on. Also, a governmental office can use multiple permissionless 
blockchains to deploy their smart contracts on. For example,
both car and house registration offices can use the Ethereum blockchain to register
cars and houses. Also, the car registration office can register some cars in the 
Ethereum blockchain while registering other cars in the Bitcoin blockchain.
Once assets are registered in a permissionless blockchains, end-users can transact
over these assets as explained next in Section~\ref{sub:proposed-transaction-model}.

\subsection{Transaction Model}\label{sub:proposed-transaction-model}

Registering complex assets in permissionless blockchains extends the 
transaction model of these blockchain. We divide the supported transactions
into \textit{four} categories as described below.

\medskip
\textbf{Currency units transactions.}
These transactions are the primitive
built-in supported transactions that allow end-users to transfer 
the ownership of currency units among end-user identities. In addition,
these transactions allow end-user to split and merge currency units as explained
in Section~\ref{sec:permissionless}.

\medskip
\textbf{Complex asset to currency units, of the same blockchain, transactions.}
These transactions allow end-users to trade complex assets 
for currency units of the same blockchain where the complex asset is registered.
These transactions are allowed through smart contract function calls. 
Smart contract classes of complex assets include the trading functionalities of 
these complex assets. For example, an end-user who wants to buy a complex asset 
calls the Buy function of the smart contract of this complex asset. This Buy 
function call is accompanied
with end-user's currency units. The Buy function transfers the currency units
to the current owner of the complex asset and transfers the ownership of the 
complex asset to the Buy function caller.
 
\medskip   
\textbf{Complex asset to currency units, of another blockchain, transactions.}
These transactions allow end-users to trade complex assets 
for currency units of a different blockchain from the one where
the complex asset is registered. These transactions are enabled
by atomic cross-chain swap protocols~\cite{herlihy2018atomic, atomicNolan, zakhary2019atomic}.
Also, these protocols require the smart contracts of complex assets
to support the functionality of atomic cross-chain transactions. For example,
an atomic cross-chain transaction could allow Alice to sell her car, registered
in the Bitcoin blockchain, to Bob who owns ether currency units in the Ethereum
blockchain. An atomic cross-chain commitment protocol must guarantee that either
both the transfer of Alice's car to Bob in the Bitcoin blockchain and the
transfer of Bob's ether to Alice in the Ethereum blockchain take place or
none of these two transfers takes place.

\medskip 
\textbf{Complex asset to complex asset transactions.}
These transactions
allow end-users to swap complex assets within the same permissionless blockchain
or across permissionless blockchains. For example, Alice might want to exchange
her car, registered in the Bitcoin blockchain, with Bob's boat, registered in
the Ethereum blockchain. These transactions use atomic cross-chain
swap protocols and require the smart contracts of complex assets
to support the functionality of atomic cross-chain transactions.

\subsection{Car Smart Contract Example}\label{sub:smart-contract-example}

\begin{algorithm}[t]
\caption{Authenticated smart contract that represents a car as a complex asset}
 \label{algo:car-smart-contract-auth}

{\sf class} CarSmartContract \{
    
\begin{algorithmic}[1]
    \State {\sf String} make \Comment{{\tt the make of the car}}
    \label{line:car-make-auth}
    \State {\sf String} model \Comment{{\tt the model of the car}}
    \State {\sf Integer} year \Comment{{\tt the manufacture year of the car}}
    \State {\sf Double} p \Comment{{\tt the price of the car (currency units)}}
    \State {\sf Address} o \Comment{{\tt the public key of the car owner}}
    \label{line:car-owner-auth}
    \State {\sf Address} co \Comment{{\tt the contract owner represented by a multisignature address of the validators}}
    \label{line:contract-owner-auth}
    \State {\sf Double} tp \Comment{{\tt the sales tax percentage}}
    \label{line:car-tp-auth}

    \Procedure{Constructor}{{\sf String} make, {\sf String} model, {\sf Integer} year, {\sf Double} p, {\sf Double} tp, {\sf Address} o} \label{line:car-constructor-auth}
        \State {\sf this}.make = make
        \State {\sf this}.model = model
        \State {\sf this}.year = year
        \State {\sf this}.p = p
        \State {\sf this}.o = o \label{line:car-owner-address-auth}
        \State {\sf this}.tp = tp
        \State {\sf this}.co = $msg.sender$ \label{line:contract-owner-address-auth}
    \EndProcedure
    
    \Procedure{Buy}{Address curOwner} \label{line:car-buy-auth}
        \State {\sf requires}($msg.val \ge  this.p \cdot (1 + \frac{this.tp}{100})$ and curOwner == this.o) \label{line:price-requirement-auth}
        \State {\sf transfer} $msg.val \cdot (1 - \frac{this.tp}{100})$  to {\sf this}.o
        \State {\sf transfer} $msg.val \cdot  \frac{this.tp}{100}$ to {\sf this}.co
        \State {\sf this}.o = $msg.sender$
    \EndProcedure
    
    \Procedure{UpdatePrice}{{\sf Double} p} \label{line:car-update-price-auth}
        \State {\sf requires}($msg.sender == $ {\sf this}.o)
        \State {\sf this}.p = p
    \EndProcedure
    
    \Procedure{UpdateContractOwner}{{\sf Address} co} \label{line:car-update-co-auth}
        \State {\sf requires}(validate-multisig($msg.sender$, {\sf this}.co))
        \State {\sf this}.co = co
    \EndProcedure
    
    \Procedure{DestroyContract}{} \label{line:contract-destroy-auth}
        \State {\sf requires}(validate-multisig($msg.sender$, {\sf this}.co))
        \State destruct-contract()
    \EndProcedure
\end{algorithmic}
\}
\end{algorithm}

This section presents an authenticated smart contract example that represents
a car as a complex asset. This example 
illustrates the necessary updates to the smart contract presented in
Algorithm~\ref{algo:car-smart-contract} in order to ensure the \textit{authenticity} and 
the \textit{legality} of the car registration in a permissionless blockchain. 
These updates are reflected in Algorithm~\ref{algo:car-smart-contract-auth}.

The member variable (Lines~\ref{line:car-make-auth}
--~\ref{line:car-tp-auth}) represent the attributes of the car. As shown, the smart contract
itself is owned by the validators multi-signature address (Line~\ref{line:contract-owner-auth}).
In addition, the car itself, as a complex asset example, is owned by an end-user
(Line~\ref{line:car-owner-auth}). 

The constructor (Line~\ref{line:car-constructor-auth}) initializes the car object 
with the attribute values passed in the deployment message (e.g., {\sf make}, {\sf model}, {\sf year}, 
{\sf price}, {\sf tax percentage}, and the {\sf owner's public key}). In addition, the constructor uses 
the implicit parameter $msg.sender$ to initialize the smart contract owner (Line~\ref{line:contract-owner-address-auth}). 

The smart contract has two functions to manipulate the car asset: 
{\sf Buy} (Line~\ref{line:car-buy-auth}) and {\sf UpdatePrice} (Line~\ref{line:car-update-price-auth}).
In addition, the smart contract has two functions to manipulate the smart contract
itself: {\sf UpdateContractOwner} (Line~\ref{line:car-update-co-auth}) and {\sf DestroyContract}
(Line~\ref{line:contract-destroy-auth}).

The {\sf Buy} function is slightly different
from the {\sf Buy} function of the smart contract in Algorithm~\ref{algo:car-smart-contract}. The 
main difference is that the validators of the permissioned blockchain embed
the taxation law in the code of the {\sf Buy} function. When a {\sf Buy} function call
is received by the permissionless blockchain mining nodes, they verify that value of
the currency units sent in $msg.val$ is greater than or equal to the sum of the 
car price and the sales tax value of this car. If true, the car price is sent to the 
current car owner, the tax value is sent to the 
contract owner (the validators' multi-signature address), and the ownership of the car is transferred to $msg.sender$.
The {\sf UpdatePrice}
function is the same as the {\sf UpdatePrice} function in 
Algorithm~\ref{algo:car-smart-contract}.

The {\sf UpdateContractOwner} function allows the validators to change the ownership of the contract
to another multi-signature address. This function is necessary to alter the contract
ownership in case a validator's identity is stolen. Validators replace the current 
multi-signature address that includes a stolen identity by a newly generated 
multi-signature address that excludes the stolen identity. The {\sf DestroyContract} function
allows the validators to destroy the smart contract object. 

\subsection{Alternative Asset Management Model} \label{sub:alternative-model}

The proposed global asset management system leverages a permissioned blockchain
only in the registration process of complex assets in a permissionless blockchain.
Once an asset is registered, the permissionless blockchain has the only record of
the current ownership of the asset in the asset's
smart contract object. In addition, the permissionless blockchain is the 
only marketplace where this asset is traded. Of course, the asset can be traded for
other assets and currency units in other permissionless blockchain through atomic
swaps. However, the asset object indefinitely remains in the same 
permissionless blockchain from its registration time until the asset's smart 
contract object is explicitly destroyed.

An alternative model can unify permissioned and permissionless blockchains as 
follows. A permissioned blockchain maintains the ownership record of an asset. If
the asset owner wants to trade it for some assets or currency units of some 
permissionless blockchain, the owner requests to register this asset in a trading
smart contract in the permissionless blockchain. The permissionless blockchain
only acts as the marketplace to trade assets. After registering the asset, end-users
can complete the trade through single-chain or cross-chain transactions. Once the
trade is completed, the ownership is updated in the permissioned blockchain
and the trading smart contract object is destroyed. This model separates the
ownership storing platform, the permissioned blockchain, from the trading platform,
the permissionless blockchain.
\section{Challenges}\label{sec:challenges}

Our proposal of unifying permissioned and permissionless blockchains to create a 
global asset 
management system faces many challenges. First, the \textit{scalability} of the
global asset management system is bounded by the scalability of the underlying 
permissionless
blockchain. Current permissionless blockchains are not scalable (e.g., Bitcoin
blockchain 
processes 3$\sim$7 transactions per second~\cite{maiyya2018database}). As a 
result, the scalability of the global
asset management system could be limited. We address the scalability 
challenge in
Section~\ref{sub:scalability-challenge}. The second challenge is 
\textit{validator identity theft}. If the identity of some
validators of the permissioned blockchain are stolen, the stolen identities
can be used to destroy currently deployed smart contracts in addition to 
authenticating smart contracts of assets that do not exist. The problem
of validator identity theft is addressed in 
Section~\ref{sub:validator-identity-theft}. Finally, we address the
\textit{asset registration flexibility} challenge. Our current model allows a 
complex asset to be registered in only one permissionless blockchain at a time. 
We discuss open research challenges that arise from allowing a complex asset
to be concurrently registered and marketed at several permissionless blockchains
in Section~\ref{sub:flexibility}.

\subsection{The Scalability Challenge} ~\label{sub:scalability-challenge}

The global asset management system requires the governmental offices
to register assets in permissionless blockchains through smart contract
deployment. Registered assets are traded through smart contract function calls
that result in transactions in the underlying permissionless blockchain.
Although, the scalability, represented by the number of executed transactions per 
second (TPS), of every individual permissionless blockchain is limited, 
the global asset management system can scale. Each governmental office
can use multiple permissionless blockchains to register different end-user assets.
Therefore, the scalability of the asset management system is not bounded by the 
scalability of an individual permissionless blockchain. Instead, the TPS of the asset 
management system can scale up to the aggregated TPS of all the permissionless 
blockchains used in registering the assets. For example, if the Bitcoin blockchain
executes up to 7 TPS and the Ethereum blockchain executes up to 25 TPS, an asset
management system that register assets in both Bitcoin and Ethereun blockchains
can scale up to 32 TPS. Using additional permissionless networks to register
assets increases the overall TPS of the asset management system.

Other permissionless blockchain scaling techniques can be used to scale
the global asset management system. One technique is 
\textit{sharding}~\cite{onsharding2018}. A permissionless blockchain is partitioned
into multiple shards and each shard is maintained by some mining nodes.
Transactions that span one shard are handled
by the mining nodes of this shard. Transactions that span multiple shards 
are handled by the mining nodes of these multiple shards and coordinated 
by an atomic swap protocol~\cite{atomicNolan,herlihy2018atomic,atomicTrading,zakhary2019atomic}.
Another technique to scale permissionless blockchains is off-chain transactions.
Lightning networks~\cite{poon2016bitcoin} can be used to execute complex assets
to currency transactions. The question of how to ensure the correctness
and tolerate maliciousness in off-chain complex assets to currency transactions
remains an open research question.

\subsection{Validator Identity Theft Challenge} ~\label{sub:validator-identity-theft}

 One of the main challenges of unifying permissioned and permissionless blockchains
is \textit{trust}. The design of Permissionless blockchains is trust-free and they
only assume that some percentage of the computing power (51\% for Bitcoin) or 
of the stake owners are honest and correct. On the other hand, permissioned 
blockchains depend on a known set of trusted identities, the validators. 
If the validator failure model include byzantine failures, typically the 
number of validators is set to $3f+1$ where $f$ is the number of validators that
can maliciously fail. For example, a permissioned blockchain with four validators
can tolerate a malicious failure of one validator. Trusting the validators of
the permissioned blockchain is necessary to trust the authentication of
smart contract that represent complex assets in the permissionless blockchain.
However, the identity theft of more than $f$ validators could result in
authenticating the registration of non existing assets and destroying the smart
contract objects of existing assets.

\begin{figure}[ht!]
	\centering
     \includegraphics[width=\columnwidth]{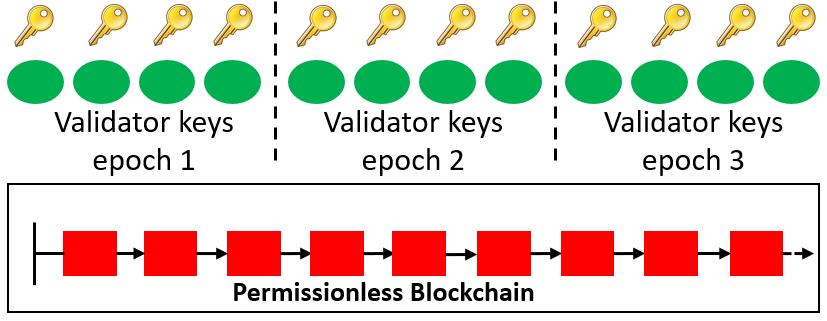}
    \caption{Permissioned blockchains use validators key rotation to limit
    the damage that results from validator identity theft.}
    \label{fig:key-rotation}
\end{figure}

To address this challenge, the standard technique of key rotation~\cite{metzger2007key, 
katar2015network} can be used to limit the damage that results from validator 
identity theft. As shown in Figure~\ref{fig:key-rotation}, the permissioned
network divides the timeline into epochs. For every epoch, a fresh set of 
validator identities is used to authenticate the smart contracts that
register assets in the permissionless blockchain during this epoch. 
A stolen validator identity set can maliciously register non existing assets 
only during one epoch. If a validator identity theft is detected within an epoch, the
permissioned blockchain can immediately reset the epoch invalidating the stolen 
validator identity. The question of how to solve the trust problem while achieving 
authenticity of asset registration remains an open research question.

\subsection{Asset Registration Flexibility Challenge} ~\label{sub:flexibility}

The proposed global asset management system requires to limit the number
of permissionless blockchain where an asset is registered to one at a time.
This requirement is necessary to prevent the \textit{double spending} of the one
asset on different blockchains.
An asset can be registered on one permissionless blockchain and if the current
asset owner wants to change the registration blockchain, an owner has to request
a contract cancellation from the validators of the permissioned blockchain. After
the contract is cancelled, the owner needs to request the registration of the 
asset in another permissionless blockchain. Asset owners might want to market 
their assets on many permissionless blockchains at a time. If this flexibility 
is allowed, a protocol is required to ensure that once the asset is traded
in a smart contract on one blockchain, other smart contracts on other blockchains
must be atomically invalidated. This protocol can be thought of as
a variation of atomic cross-chain swaps. The design of such protocol and the details
of its correctness remain open research questions.
\section{Related Work}\label{sec:related-work}

Bitcoin~\cite{nakamoto2008bitcoin} is considered the first successful global scale 
peer-to-peer cryptocurrency. The Nakamoto consensus protocol used in Bitcoin allows participants to transact with each other without 
the need for a trusted third party, such as a banks or a credit card company. 
The ledger that records all the transaction history in traditional 
trusted banks is replaced by a distributed ledger stored in all the 
participants in Bitcoin, thus eliminating
the need for trusted third parties.
Many of the recent works on permissionless blockchains are focused on enhancing one aspect of Bitcoin -- the performance limitation.
BitcoinNG~\cite{eyal2016bitcoin} separates the blocks in the chain into \textit{key-blocks} and
\textit{micro-blocks}. Key-blocks are created by solving
the proof-of-work challenge and the miner who solved the 
puzzle becomes the \textit{leader} producing many micro-blocks
consisting of transactions. The leader is replaced when another
miner mines the next key-block. Thus, by increasing the frequency of
micro-blocks produced by a leader, BitcoinNG improves the throughput
of Bitcoin. But empowering a single leader to produce micro-blocks
entails considerable risks.
ByzCoin~\cite{kogias2016enhancing} identifies the benefits of
separating the blocks into key and micro blocks, as well as the
issues with a single leader. ByzCoin replaces a single leader
with a dynamically changing group of \textit{trustees}. Trustees
execute PBFT~\cite{castro1999practical} to decide on the next
micro-block and use Collective Signing (CoSi)~\cite{syta2016keeping}
to collectively sign the chosen block.
Elastico~\cite{luu2016secure} is another blockchain solution
aiming to increase the performance of Bitcoin. The key idea
proposed in Elastico is to split all the servers in the system
into smaller sized groups called \textit{committees}. Every
committee is then assigned with a disjoint set of transactions
and the committee members verify those transactions. Each
committee executes classical PBFT in order to agree on the
set of verified transactions. These transactions are then sent
to a \textit{final committee} which is in-charge of aggregating
all the transactions produced by different committees into one
block and then to broadcast the final block. Thus by allowing
different committees to process different {\em shard} of
transactions, Elastico increases the throughput of Bitcoin.

Although the above discussed solutions provide various strategies 
to increase the performance on Bitcoin, most of the solutions
assume a cryptocurrency application. Even if they can be easily
extended to include smart contracts, they would still lack in 
managing global assets. The high churn of participants in a
permissionless blockchain network poses impediments is regulating
laws associated with global assets. 

While Permissionless blockchains are public and anyone can participate
without a specific identity, in permissioned blockchains nodes are known and identified.
In this paper, permissioned blockchains are used to register global assets and deploy them on
permissionless blockchains.
Most existing permissioned blockchains follow the order-execute architecture where
nodes agree on the order of incoming requests and then execute the requests in the same order.
Permissioned blockchains differ mainly in their consensus protocols.
Tendermint \cite{kwon2014tendermint} is different from the original PBFT in two ways,
first, only a subset of nodes, called validators, participate in the consensus protocol and second,
the leader is changed after the construction of every block (leader rotation).
Quorum \cite{morgan2016quorum} is an enterprise-focused version of Ethereum \cite{wood2014ethereum} developed by JP Morgan.
Quorum introduces a consensus protocol based on Raft \cite{ongaro2014search}: a well-known crash fault-tolerant protocol.
Chain Core \cite{chain}, Multichain \cite{greenspan2015multichain},
Hyperledger Iroha \cite{iroha}, and Corda \cite{corda} are some other prominent permissioned blockchains that
follow order-execute architecture and use varients of Byzantine fault-tolerant protocols.

Fabric \cite{androulaki2018hyperledger} introduces the execute-order-validate architecture and
leverages parallelism by executing the transactions of different applications simultaneously.
Modular design, pluggable fault-tolerant protocol,
policy-based endorsement, and non-deterministic execution
are some of the main advantages of Fabric.
However, it performs poorly on workloads with high-contention, i.e., many {\em conflicting transactions} in a block.
To support conflicting transactions, Parblockchain \cite{amiri2019parblockchain} 
introduces the order-(parallel)execute architecture where the orderer nodes
generate a dependency graph in the ordering phase and transactions are executed in parallel following
the generated dependency graph in the execution phase.

Users on the same or different blockchains should be able to initiate transactions in order to exchange assets.
Our proposal supports four types of transactions:
transactions in currency units,
transactions between complex asset and currency units in the same blockchain,
transactions between complex asset and currency units in different blockchains, and
transactions between two complex assets.
Different techniques have been presented to support intra- and cross-chain asset trades.
Atomic cross-chain swaps \cite{herlihy2018atomic}
are used for trading assets on two unrelated blockchains.
Atomic swaps use hash-lock and time-lock mechanisms to either perform all or none
of a cryptographically linked set of transactions.
Interledger protocols (ILPV \cite{thomas2015protocol}) which are presented by
the World Wide Web Consortium (W3C) use a generalization of atomic swaps
and enable secure transfers between two blockchain ledgers
using escrow transactions.
since the redemption of an escrow transaction needs fulfillment of all the terms of an agreement,
the transfer is atomic.
Lightning network \cite{miller2017sprites}\cite{poon2016bitcoin} also 
generalizes atomic swap to transfer assets between two different clients via a network of micro-payment channels.
Blocknet \cite{blocknet}, BTC \cite{buterin2016chain}, Xclaim \cite{zamyatinxclaim},
POA Bridge \cite{poa} (designed specifically for Ethereum), Wanchain \cite{wanchain}, and Fusion \cite{fusion}
are some other blockchain systems that allow users to transfer assets between two chains.

Hyperledger also addresses atomic cross-chain swap between permissioned blockchains that are deployed on different channels
by either assuming the existence of a trusted channel among the participants or
using an atomic commit protocol \cite{androulaki2018channels}\cite{androulaki2018hyperledger}.

Using sidechain is proposed in \cite{back2014enabling} 
to transfer assets from a main blockchain to the sidechain(s)
and execute some transactions in the sidechain(s) in order to
reduce confirmation time and transaction cost, and 
support more functionality.
Liquid \cite{dilley2016strong}, Plasma \cite{poon2017plasma},
Sidechains \cite{garoffolo2018sidechains}, and RSK \cite{lerner2015rootstock}
are some other blockchain systems that use sidechains.
In Polkadot \cite{wood2016polkadot} and Cosmos \cite{kwon2016cosmos} also assets can be exchanged
using a main chain and a set of (side) blockchains.
Both Polkadot and Cosmos rely on byzantine consensus protocol in both sender and receiver sides.

To support global assets in blockchains, using tokens which are backed by external assets, called asset-backed tokens,
is proposed \cite{asset-backed}.
Tokenization is the process of representing the ownership of real world assets digitally on a blockchain.
While the main purpose of tokenization is to use tokens as assets (investment instrument) and split it into smaller pieces,
in this paper we mainly focus on how to authenticate an asset as being legitimate
so that it can be transacted in a marketplace (i.e., transfer of ownership).
In addition, the tokenization of the assets on the blockchain is being done by a known entity (highly centralized)
whereas in our proposal the centralized entity is replaced by a governmental permissioned blockchain
which first, puts the responsibility of forcing the law on the government, and
second, ensures that the centralized entities do not monopolize the tokenization of assets on the blockchain.
\section{Conclusion}\label{sec:conclusion}

In this paper, we propose a global asset management system that leverages
both permissioned and permissionless blockchains. Governmental offices maintain
trusted permissioned blockchains. Permissioned blockchains \textit{authenticate}
the registration of end-user assets in permissionless blockchains through smart 
contracts. In addition, permissioned blockchains prevent the \textit{double spending}
of assets by ensuring that every asset can be registered in only
one authenticated smart contract in one permissionless blockchain. Finally, the
permissioned blockchain ensure the \textit{legality} of trading the assets by
encoding the laws (e.g., taxation law) in the smart contract code. Permissionless
blockchains are marketplaces to trade the registered assets. Registered assets
can be traded for currency units or other assets on the same permissionless blockchain
or on other permissionless blockchain. This extended transaction model is enabled 
through single-chain and cross-chain transactions.

\balance
\bibliographystyle{abbrv}
\bibliography{main}

\begin{thebibliography}{10}

\bibitem{chain}
Chain.
\newblock {http://chain.com}.

\bibitem{corda}
Corda.
\newblock {https://github.com/corda/corda}.

\bibitem{fusion}
Fusion whitepaper: An inclusive cryptofinance platform based on blockchain.
\newblock {https://docs.wixstatic.com/ugd/
  76b9ac\_be5c61ff0e3048b3a21456223d542687.pdf}.

\bibitem{iroha}
Hyperledger iroha.
\newblock {https://github.com/hyperledger/iroha}.

\bibitem{poa}
Poa bridge.
\newblock {https://github.com/poanetwork/token-bridge}.

\bibitem{wanchain}
Wanchain: Building super financial markets for the new digital economy.
\newblock {https://www.wanchain.org/files/Wanchain-Whitepaper-EN-version.pdf}.

\bibitem{atomicTrading}
Atomic cross-chain trading.
\newblock \url{https://en.bitcoin.it/wiki/Atomic_cross-chain_trading}, 2018.

\bibitem{solidity}
Solidity — solidity 0.5.5 documentation.
\newblock \url{https://solidity.readthedocs.io/en/v0.5.5/}, 2018.

\bibitem{amiri2019parblockchain}
M.~J. Amiri, D.~Agrawal, and A.~E. Abbadi.
\newblock Parblockchain: Leveraging transaction parallelism in permissioned
  blockchain systems.
\newblock In {\em 2019 IEEE 39th International Conference on Distributed
  Computing Systems (ICDCS)}. IEEE, 2019.

\bibitem{androulaki2018hyperledger}
E.~Androulaki, A.~Barger, V.~Bortnikov, C.~Cachin, K.~Christidis, A.~De~Caro,
  D.~Enyeart, C.~Ferris, G.~Laventman, Y.~Manevich, et~al.
\newblock Hyperledger fabric: a distributed operating system for permissioned
  blockchains.
\newblock In {\em Proceedings of the Thirteenth EuroSys Conference}, page~30.
  ACM, 2018.

\bibitem{androulaki2018channels}
E.~Androulaki, C.~Cachin, A.~De~Caro, and E.~Kokoris-Kogias.
\newblock Channels: Horizontal scaling and confidentiality on permissioned
  blockchains.
\newblock In {\em European Symposium on Research in Computer Security}, pages
  111--131. Springer, 2018.

\bibitem{back2014enabling}
A.~Back, M.~Corallo, L.~Dashjr, M.~Friedenbach, G.~Maxwell, A.~Miller,
  A.~Poelstra, J.~Tim{\'o}n, and P.~Wuille.
\newblock Enabling blockchain innovations with pegged sidechains.
\newblock 2014.

\bibitem{buterin2016chain}
V.~Buterin.
\newblock Chain interoperability.
\newblock {\em R3 Research Paper}, 2016.

\bibitem{onsharding2018}
V.~Buterin.
\newblock On sharding blockchains.
\newblock \url{https://github.com/ethereum/wiki/wiki/Sharding-FAQs}, 2018.

\bibitem{buterin2014next}
V.~Buterin et~al.
\newblock A next-generation smart contract and decentralized application
  platform.
\newblock {\em white paper}, 2014.

\bibitem{cachin2016architecture}
C.~Cachin.
\newblock Architecture of the hyperledger blockchain fabric.
\newblock In {\em Workshop on Distributed Cryptocurrencies and Consensus
  Ledgers}, volume 310, 2016.

\bibitem{cachin2017blockchain}
C.~Cachin and M.~Vukoli{\'c}.
\newblock Blockchain consensus protocols in the wild.
\newblock In {\em 31 International Symposium on Distributed Computing, {DISC}},
  pages 1--16, 2017.

\bibitem{castro1999practical}
M.~Castro, B.~Liskov, et~al.
\newblock Practical byzantine fault tolerance.
\newblock In {\em OSDI}, volume~99, pages 173--186, 1999.

\bibitem{morgan2016quorum}
J.~M. Chase.
\newblock Quorum white paper, 2016.

\bibitem{blocknet}
A.~Culwick and D.~Metcalf.
\newblock The blocknet design specification, 2018.

\bibitem{dickerson2017adding}
T.~Dickerson, P.~Gazzillo, M.~Herlihy, and E.~Koskinen.
\newblock Adding concurrency to smart contracts.
\newblock In {\em Proceedings of the ACM Symposium on Principles of Distributed
  Computing}, pages 303--312. ACM, 2017.

\bibitem{dilley2016strong}
J.~Dilley, A.~Poelstra, J.~Wilkins, M.~Piekarska, B.~Gorlick, and
  M.~Friedenbach.
\newblock Strong federations: An interoperable blockchain solution to
  centralized third-party risks.
\newblock {\em arXiv preprint arXiv:1612.05491}, 2016.

\bibitem{eyal2016bitcoin}
I.~Eyal, A.~E. Gencer, E.~G. Sirer, and R.~Van~Renesse.
\newblock Bitcoin-ng: A scalable blockchain protocol.
\newblock In {\em NSDI}, pages 45--59, 2016.

\bibitem{garoffolo2018sidechains}
A.~Garoffolo and R.~Viglione.
\newblock Sidechains: Decoupled consensus between chains.
\newblock {\em arXiv preprint arXiv:1812.05441}, 2018.

\bibitem{greenspan2015multichain}
G.~Greenspan.
\newblock Multichain private blockchain-white paper.
\newblock {\em URl: http://www. multichain.
  com/download/MultiChain-White-Paper. pdf}, 2015.

\bibitem{herlihy2018atomic}
M.~Herlihy.
\newblock Atomic cross-chain swaps.
\newblock In {\em ACM Symposium on Principles of Distributed Computing (PODC)},
  pages 245--254. ACM, 2018.

\bibitem{herlihy2019blockchains}
M.~Herlihy.
\newblock Blockchains from a distributed computing perspective.
\newblock {\em Communications of the ACM}, 62(2):78--85, 2019.

\bibitem{asset-backed}
E.~Hill.
\newblock What is an asset-backed token? a complete guide to security token
  assets.
\newblock
  \url{https://medium.com/ico-launch-malta/what-is-an-asset-backed-token-a-complete-guide-to-security-token-assets-f7a0f111d443}.

\bibitem{katar2015network}
S.~Katar, L.~W. Yonge, and M.~Krishnam.
\newblock Network encryption key rotation, Mar.~24 2015.
\newblock US Patent 8,989,379.

\bibitem{kogias2016enhancing}
E.~K. Kogias, P.~Jovanovic, N.~Gailly, I.~Khoffi, L.~Gasser, and B.~Ford.
\newblock Enhancing bitcoin security and performance with strong consistency
  via collective signing.
\newblock In {\em 25th USENIX Security Symposium (USENIX Security 16)}, pages
  279--296, 2016.

\bibitem{kwon2014tendermint}
J.~Kwon.
\newblock Tendermint: Consensus without mining.
\newblock {\em Draft v. 0.6, fall}, 2014.

\bibitem{kwon2016cosmos}
J.~Kwon and E.~Buchman.
\newblock Cosmos: A network of distributed ledgers.
\newblock {\em URL https://cosmos. network/whitepaper}, 2016.

\bibitem{lamport1978time}
L.~Lamport.
\newblock Time, clocks, and the ordering of events in a distributed system.
\newblock {\em Communications of the ACM}, 21(7):558--565, 1978.

\bibitem{lamport2001paxos}
L.~Lamport et~al.
\newblock Paxos made simple.
\newblock {\em ACM Sigact News}, 32(4):18--25, 2001.

\bibitem{lerner2015rootstock}
S.~D. Lerner.
\newblock Rootstock: Bitcoin powered smart contracts, 2015.

\bibitem{luu2016secure}
L.~Luu, V.~Narayanan, C.~Zheng, K.~Baweja, S.~Gilbert, and P.~Saxena.
\newblock A secure sharding protocol for open blockchains.
\newblock In {\em Proceedings of the 2016 ACM SIGSAC Conference on Computer and
  Communications Security}, pages 17--30. ACM, 2016.

\bibitem{maiyya2018database}
S.~Maiyya, V.~Zakhary, D.~Agrawal, and A.~E. Abbadi.
\newblock Database and distributed computing fundamentals for scalable,
  fault-tolerant, and consistent maintenance of blockchains.
\newblock {\em Proceedings of the VLDB Endowment}, 11(12):2098--2101, 2018.

\bibitem{metzger2007key}
B.~Metzger, S.~Mauldin, B.~Sandell, and J.~Chang.
\newblock Key rotation, Mar.~29 2007.
\newblock US Patent App. 11/236,046.

\bibitem{miller2017sprites}
A.~Miller, I.~Bentov, R.~Kumaresan, C.~Cordi, and P.~McCorry.
\newblock Sprites and state channels: Payment networks that go faster than
  lightning.
\newblock {\em arXiv preprint arXiv:1702.05812}, 2017.

\bibitem{mohan2017tutorial}
C.~Mohan.
\newblock Tutorial: blockchains and databases.
\newblock {\em Proceedings of the VLDB Endowment}, 10(12):2000--2001, 2017.

\bibitem{nakamoto2008bitcoin}
S.~Nakamoto.
\newblock Bitcoin: A peer-to-peer electronic cash system.
\newblock 2008.

\bibitem{atomicNolan}
T.~Nolan.
\newblock Alt chains and atomic transfers.
\newblock
  \url{https://bitcointalk.org/index.php?topic=193281.msg2224949#msg2224949},
  2013.

\bibitem{ongaro2014search}
D.~Ongaro and J.~K. Ousterhout.
\newblock In search of an understandable consensus algorithm.
\newblock In {\em USENIX Annual Technical Conference}, pages 305--319, 2014.

\bibitem{poon2017plasma}
J.~Poon and V.~Buterin.
\newblock Plasma: Scalable autonomous smart contracts.
\newblock {\em White paper}, 2017.

\bibitem{poon2016bitcoin}
J.~Poon and T.~Dryja.
\newblock The bitcoin lightning network: Scalable off-chain instant payments,
  2016.

\bibitem{rivest1978method}
R.~L. Rivest, A.~Shamir, and L.~Adleman.
\newblock A method for obtaining digital signatures and public-key
  cryptosystems.
\newblock {\em Communications of the ACM}, 21(2):120--126, 1978.

\bibitem{syta2016keeping}
E.~Syta, I.~Tamas, D.~Visher, D.~I. Wolinsky, P.~Jovanovic, L.~Gasser,
  N.~Gailly, I.~Khoffi, and B.~Ford.
\newblock Keeping authorities" honest or bust" with decentralized witness
  cosigning.
\newblock In {\em 2016 IEEE Symposium on Security and Privacy (SP)}, pages
  526--545. Ieee, 2016.

\bibitem{thomas2015protocol}
S.~Thomas and E.~Schwartz.
\newblock A protocol for interledger payments.
\newblock 2015.

\bibitem{wood2014ethereum}
G.~Wood.
\newblock Ethereum: A secure decentralised generalised transaction ledger.
\newblock {\em Ethereum project yellow paper}, 151:1--32, 2014.

\bibitem{wood2016polkadot}
G.~Wood.
\newblock Polkadot: Vision for a heterogeneous multi-chain framework.
\newblock {\em White Paper}, 2016.

\bibitem{zakhary2019atomic}
V.~Zakhary, D.~Agrawal, and A.~E. Abbadi.
\newblock Atomic commitment across blockchains.
\newblock {\em arXiv preprint arXiv:1905.02847}, 2019.

\bibitem{zamyatinxclaim}
A.~Zamyatin, D.~Harz, J.~Lind, P.~Panayiotou, A.~Gervais, and W.~J.
  Knottenbelt.
\newblock Xclaim: A framework for blockchain interoperability.
\newblock 2019.

\end{thebibliography}

\end{document}